\documentclass[structabstract]{aa}
\usepackage{txfonts}
\usepackage{graphicx}      
\usepackage{url}
\usepackage[squaren,cdot]{SIunits} % SI units of measurement (\centi\meter)
\usepackage{xspace}                % To let a space in the command \he
\usepackage{natbib}
\usepackage{multirow}      

\newcommand{\um}{\object{UM673}\xspace } % xspace lets a space, but not in presence of punctuation
\newcommand{\he}{\object{HE 0435-1223}\xspace }

\begin{document}
\title{Flux and color variations of the doubly imaged quasar UM673
  \thanks{Based on data collected by MiNDSTEp with the Danish 1.54m
    telescope at the ESO La Silla Observatory. Light curves are
    available via http://cdsweb.u-strasbg.fr/cgi-bin/qcat?J/A+A/}}

% \subtitle{}
\author{
  D.~Ricci\inst{1}\fnmsep\inst{30}\fnmsep\inst{31} \and
  A.~Elyiv\inst{1}\fnmsep\inst{2}\and
  F.~Finet\inst{1}\and
  O.~Wertz\inst{1}\and
  K.~Alsubai\inst{27}\and
  T.~Anguita\inst{3}\fnmsep\inst{4}\and 
  V.~Bozza\inst{5}\fnmsep\inst{6}\and 
  P.~Browne\inst{7}\and 
  M.~Burgdorf\inst{8}\fnmsep\inst{24}\and
  S.~Calchi Novati\inst{5}\fnmsep\inst{9}\and 
  P.~Dodds\inst{7}\and 
  M.~Dominik\inst{7}\fnmsep\thanks{Royal Society University Research Fellow}\and 
  S.~Dreizler\inst{10}\and
  T.~Gerner\inst{15}\and 
  M.~Glitrup\inst{11}\and 
  F.~Grundahl\inst{11}\and 
  S.~Hardis\inst{12}\and 
  K.~Harps\o e\inst{12}\fnmsep\inst{25}\and 
%  F.~Hessman\inst{10}\and 
  T.~C.~Hinse\inst{12}\fnmsep\inst{13}\and 
  A.~Hornstrup\inst{14}\and 
  M.~Hundertmark\inst{10}\and 
  U.~G.~J\o rgensen\inst{12}\fnmsep\inst{25}\and 
  N.~Kains\inst{7}\and 
  E.~Kerins\inst{26}\and 
  C.~Liebig\inst{7}\fnmsep\inst{15}\and 
  G.~Maier\inst{15}\and 
  L.~Mancini\inst{4}\fnmsep\inst{5}\and
  G.~Masi\inst{17}\and 
  M.~Mathiasen\inst{12}\and 
  M.~Penny\inst{26}\and
  S.~Proft\inst{15}\and 
  S.~Rahvar\inst{18}\fnmsep\inst{29}\and 
  G.~Scarpetta\inst{5}\fnmsep\inst{6}\and 
  K.~Sahu\inst{28}\and
  S.~Sch\"afer\inst{10}\and
  F.~Sch\"onebeck\inst{15}\and
  R.~Schmidt\inst{15}\and
  J.~Skottfelt\inst{12}\and 
  C.~Snodgrass\inst{19}\fnmsep\inst{20}\and 
  J.~Southworth \inst{21}\and 
%  J.~Teuber\inst{12}\and 
  C.~C.~Th\"one\inst{22}\fnmsep\inst{23}\and 
  J.~Wambsganss\inst{15}\and 
  F.~Zimmer\inst{15}\and 
  M.~Zub\inst{15} 
  \and
  J.~Surdej\inst{1}\fnmsep\thanks{also Directeur de Recherche honoraire du FRS-FNRS}
}
\institute{%1  
  D\'epartement d'Astrophysique, G\'eophysique et Oc\'eanographie, B\^at.~B5C, Sart Tilman,
  Universit\'e de  Li\`ege, 
  4000 Li\`ege 1, Belgique;
  \email{ricci@astro.ulg.ac.be}
  \and %2
  Main Astronomical Observatory, Academy of Sciences of Ukraine, Zabolotnoho 27, 03680 Kyiv, Ukraine
  \and %3
  Centro de Astro-Ingenier\'ia, Departamento de Astronom\'ia y Astrof\'isica, Pontificia Universidad Cat\'olica de Chile, Casilla 306, Santiago, Chile.
  \and %4
  Max-Planck-Institut f\"ur Astronomie, K\"onigstuhl 17, 69117 Heidelberg, Germany
  \and %5
  Dipartimento di Fisica ``E.R. Caianiello'', Universit\`a degli Studi di Salerno, Via Ponte Don Melillo, 84085 Fisciano (SA), Italy
  \and %6
  Instituto Nazionale di Fisica Nucleare, Sezione di Napoli, Italy
  \and %7
  SUPA, University of St~Andrews, School of Physics \& Astronomy, North Haugh, St~Andrews, KY16 9SS, United Kingdom
  \and %8
  Deutsches SOFIA Institut, Universitaet Stuttgart, Pfaffenwaldring 31, 70569 Stuttgart, Germany
  \and %9
  Istituto Internazionale per gli Alti Studi Scientifici (IIASS), Vietri Sul Mare (SA), Italy
  \and %10
  Institut f\"ur Astrophysik, Georg-August-Universit\"at G\"ottingen, Friedrich-Hund-Platz 1, 37077 G\"ottingen, Germany
  \and %11
  Department of Physics \& Astronomy, Aarhus University, Ny Munkegade, 8000 Aarhus C, Denmark
  \and %12
  Niels Bohr Institute, University of Copenhagen, Juliane Maries vej 30, 2100 Copenhagen \O, Denmark
  \and %13
  KASI - Korea Astronomy and Space Science Institute, 776 Daedukdae-ro, Yuseong-gu, Daejeon 305-348, Republic of Korea
  \and %14
  National Space Institute, Technical University of Denmark, 2800 Lyngby, Denmark
  \and %15
  Astronomisches Rechen-Institut, Zentrum f\"ur Astronomie, Universit\"at Heidelberg, M\"onchhofstra\ss e 12-14, 69120 Heidelberg, Germany
  \and %16
  Dipartimento di Ingegneria, Universit\`a del Sannio, Corso Garibaldi 107, 82100 Benevento, Italy
  \and %17
  Bellatrix Astronomical Observatory, Center for Backyard Astrophysics, Ceccano (FR), Italy
  \and %18
  Physics Department, Sharif University of Technology, Tehran, Iran
  \and %19
  European Southern Observatory, Casilla 19001, Santiago 19, Chile
  \and %20
  Max Planck Institute for Solar System Research, Max-Planck-Str. 2, 37191 Katlenburg-Lindau, Germany
  \and %21
  Astrophysics Group, Keele University, Newcastle-under Lyme, ST5 5BG, United Kingdom
  \and %22
  Dark Cosmology Centre, Niels Bohr Institute, University of Copenhagen, Juliane Maries Vej 30, Copenhagen Ø, 2100 Denmark  
  \and %23
  INAF, Osservatorio Astronomico di Brera, 23807 Merate, Italy
  \and %24
  SOFIA Science Center, NASA Ames Research Center, Mail Stop N211-3, Moffett Field CA 94035, USA
  \and %25
  Centre for Star and Planet Formation, Geological Museum, \O ster Voldgade 5, 1350 Copenhagen, Denmark.
  \and %26
  Jodrell Bank Centre for Astrophyics, University of Manchester, United Kingdom
  \and %27
  Alsubai's Establishment for Scientific Studies, Qatar
  \and %28  
  Space Telescope Science Institute (STScI), United States of America
  \and %29
  Perimeter Institute for Theoretical Physics, 31 Caroline Street North, Waterloo, Ontario N2L 2Y5, Canada
  \and %30
  INAF/Istituto di Astrofisica Spaziale e Fisica Cosmica, Bologna, Via Gobetti 101, 40129 Bologna, Italy.
  \and %31 
  Instituto de Astronom\'ia, Universidad Nacional Aut\'onoma de M\'exico, Apdo. Postal 877, Ensenada, B.C. 22800, Mexico
}

\date{\today}

\abstract
% context heading (optional), leave it empty if necessary
    {}
    % aims heading (mandatory)   
    { With the aim of characterizing the flux and color variations of
      the multiple components of the gravitationally lensed quasar \um
      as a function of time, we have performed multi-epoch and
      multi-band photometric observations with the Danish $1.54\meter$
      telescope at the La Silla Observatory. }
    % methods heading (mandatory) 
    { The observations were carried out in the $VRi$ spectral bands
      during four seasons (2008--2011). We reduced the data using
      the PSF (Point Spread Function) photometric technique as well as aperture photometry. }
    % results heading (mandatory)
    { Our results show for the brightest lensed component some
      significant decrease in flux between the first two
          seasons ($+0.09$/$+0.11$/$+0.05$ mag) and a subsequent
          increase during the following ones
          ($-0.11$/$-0.11$/$-0.10$ mag) in the $V/R/i$
          spectral bands, respectively. Comparing our results with previous
      studies, we find smaller color variations between these seasons
      as compared with previous ones.  We also separate the
      contribution of the lensing galaxy from that of the fainter and
      close lensed component.
    }
    % conclusions heading (optional), leave it empty if necessary
    {}
    
    \maketitle
    
    %%%%%%%%%%%%%%%%%%%%%%%%%%%%%%%%%%%%%%%%%%%%%%%%%%%%%%%%%%%%%%%%%%%%% 
\section{Introduction}

Multiply imaged quasars are of great interest in astrophysics due to
the possibility, from observed flux and color variations between the
lensed components, to distinguish between intrinsic quasar variations
caused by the accretion mechanism, and microlensing effects induced by
stars in the lens galaxy \citep{wambsganss06}.

In a previous paper \citep{ricci11,ricci11-cat}, we have studied such
variations for the quadruply imaged quasar \he, observed in the
framework of a $VRi$ multi-epoch monitoring of five lensed
quasars\footnote{ UM673/\object{Q0142-100}, \object{HE0435-1223},
  \object{Q2237+0305}, \object{WFI2033-4723} and
  \object{HE0047-1756}}, a parallel project of the MiNDSTEp
(Microlensing Network for the Detection of Small Terrestrial
Exoplanets) campaign \citep{dominik10}.

\begin{figure}[tbp]
  \centering \includegraphics[width=8.7cm]{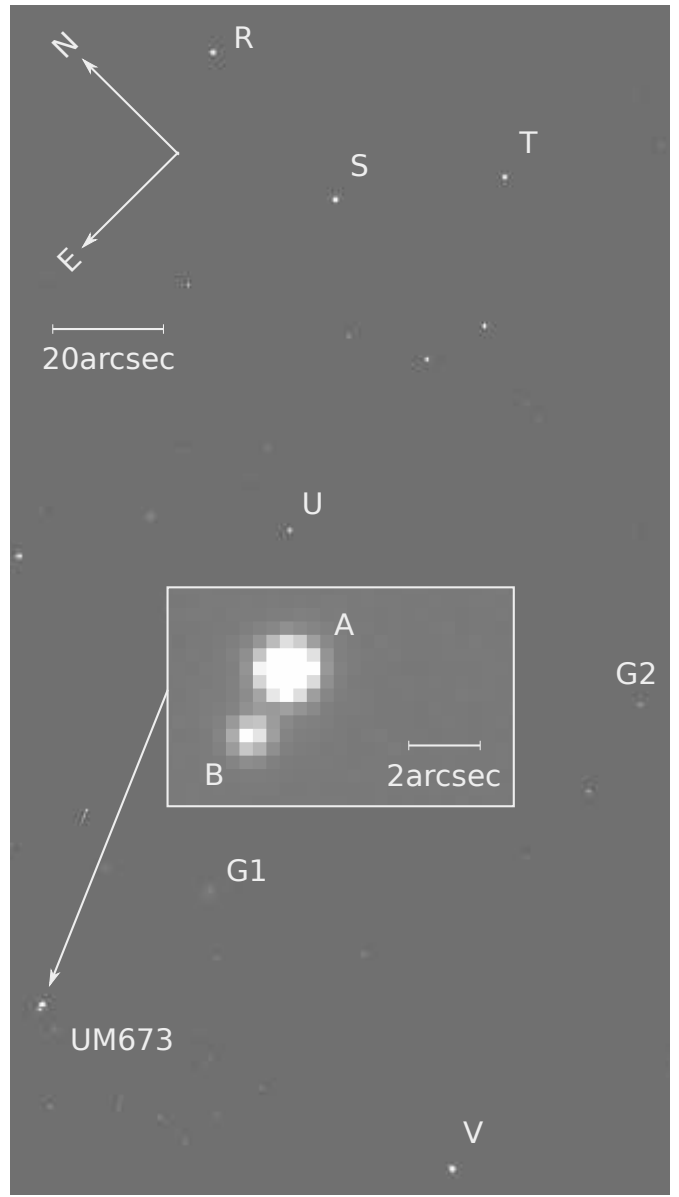}
  \caption{DFOSC $V$ filter image, taken on 2008-08-03, showing the
    position of \um and the stars ``R'', ``S'', ``T'', ``U'', and
    ``V'' used to search for a suitable reference star. The ``V'' star
    was finally chosen. ``G1'' and ``G2'' are field galaxies. The
    inset zoom shows the two components ``A'' and ``B'' of the lensed
    quasar.}
  \label{fig:full}
\end{figure}

In the current paper, we focus on \um/\object{Q0142--100} (see
Fig.~\ref{fig:full}), a doubly imaged quasar discovered by
\cite{surdej87} during a high resolution imaging survey of HLQs
(Highly Luminous Quasars) and subsequently studied by our team
\citep{smette90,smette92,daulie93,nakos05}.

\cite{surdej88} reported a separation of $2.22\arcsec$ between the
components ``A'' (brighter) and ``B'' (fainter), and found their $V$
magnitudes to be $16.9$ and $19.1$ respectively, at a redshift
$z=2.719$. The redshift of the sensibly fainter ($R=19.2$) lensing
galaxy, located very close to the ``B'' component, was derived to be
$z=0.49$, and the time delay between the two lensed components was
estimated around 7 weeks.

A photometric monitoring of \um was performed during the years
1987--1993 \citep{daulie93}, but the photometry did not show any clear
evidence for relative variations over the considered period.

In the framework of the \textsc{castles} (CfA Arizona Space Telescope
LEns Survey) project, precise astrometry of the components and of the
lens galaxy ``G'' was
obtained\footnote{\url{http://www.cfa.harvard.edu/castles/Individual/Q0142.html}}.
The colors of the lens galaxy were found to match those of a passively
evolving early-type galaxy at $z\approx0.5$ \citep{castles}.

% -------------------------------------------------------
% -------------------------------------------------------
% =====================================
\begin{table}[tbp]
  \caption{Number of CCD images collected for each filter and each year of
    observation of \um. The corresponding number of nights
    for each filter is also shown.}
\label{tab:obs}
\centering
\begin{tabular}{crrrrcrrrr}
\hline\hline
       & \multicolumn{4}{c}{images} & & \multicolumn{4}{c}{nights}  \\
% \cline{2-5} \cline{7-10}
season & $V$ & $R$ & $i$ &   total  & & $V$ & $R$ & $i$  & total\\
\hline                                                 
 2008  & 42  & 45  & 43  &   130    & & 15 & 15  & 15  & 45 \\
 2009  & 34  & 35  & 26  &    95    & & 12 & 13  &  9  & 34 \\
 2010  & 72  & 78  &  0  &   150    & & 23 & 26  &  0  & 49 \\
 2011  & 51  & 53  &  9  &   113    & & 15 & 16  &  1  & 32 \\
total  & 199 & 211 & 78  &   488    & & 65 & 70  & 25  & 160 \\
\hline                         
\end{tabular}
\end{table}
% =====================================

\cite{lehar00, lehar00-err} reported HST observations of \um at
optical and infrared wavelengths, and \cite{sinachopoulos01a} observed
the lensed quasar in the $R$ filter for six seasons (1995--2000),
detecting a significant increase by $0.3$ mag of the combined
system (lensed components) with respect to the values reported at
discovery, with a peak of $0.5$ mag during the period
1995--1997. \cite{lehar00} performed photometric measurements on HST
(Hubble Space Telescope) data taken in the $R$ filter, and obtained
magnitudes of $16.67$, $18.96$, and $19.35$ for the ``A'', ``B''
components and the lens galaxy, respectively.

After spectrophotometric observations performed in 2002 by
\cite{wisotzki04}, which did not show any evidence of microlensing,
the first multi-filter monitoring of \um was carried out by
\cite{nakos05} between 1998 and 1999, in the Cousins $V$ and Gunn $i$
filters. Analysis of the light curves was made using three different
photometric methods: image deconvolution \citep{magain07}, PSF (Point
Spread Function) fitting, and image subtraction. \cite{nakos05} found
that component ``A'' displayed possible evidence for microlensing.

  \cite{koptelova08,koptelova11,koptelova10} and \cite{koptelova10b}
  observed the object in the $VRI$ bands and succeeded for the first
time in determining a time delay: $150^{+7\ \ +42}_{-18\ -36}$ days
(at $68\%$ and $95\%$ confidence levels).

Furthermore, \cite{fadley11} examined the wavelength dependence of the
flux ratios for several gravitationally lensed quasars using $K$ and
$L'$-band images obtained with the Gemini North $8\meter$ telescope,
detecting no difference between the two flux ratios for the specific
case of \um (``B''/``A'' $= 0.128\pm 0.002$ in the $K$-band and
$0.132\pm 0.006$ in the $L'$-band).

Finally, in a recent paper, \cite{koptelova12}
    re-estimated the determination of the time delay to a value of
    $89\pm 11$ days using 2001--2011 VRI observations, and suggested
    the brightness variations to be mainly due to intrinsic variations
    of the quasar.  

We present multi-epoch photometric monitoring data of \um over four
seasons (2008--2011), carried out in three filters ($VRi$) with the
DFOSC (Danish Faint Object Spectrograph and Camera) instrument of the
Danish $1.54\meter$ telescope at the La Silla Observatory.

The observations and the pre-processing of the images are presented in
Sect.~\ref{sec:obs}. Sect.~\ref{sec:red} presents the reduction
techniques and the results are shown in Sect.~\ref{sec:res}. Finally,
Sect.~\ref{sec:conc} contains the main conclusions.

%%%%%%%%%%%%%%%%%%%%%%%%%%%%%%%%%%%%%%%%%%%%%%%%%%%%%%%%%%%%%%%%%%%%% 
\section{Observations and pre-processing}
\label{sec:obs}
% -------------------------------------------------------
\begin{figure}[tbp]
%  \centering \includegraphics[height=5.0cm]{galaxy-lens.eps}
  \centering \includegraphics[width=8.7cm]{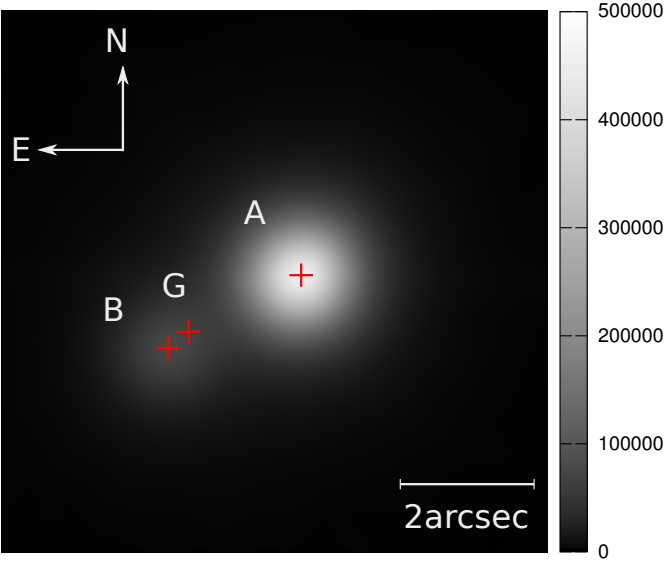}
  \caption{Composite image of \um obtained by superposing the 44 best
    quality CCD images in the $R$ filter, resampled by dividing each
    pixel in a grid of $20\times 20$ subpixels and recentering the
    images with an accuracy of one new subpixel. The positions of the
    components, provided by HST, are also shown.}
  \label{fig:galaxy-lens}
\end{figure}
% ------------------------------------------------------
% =====================================
\begin{table}[bp]
  \caption{Maximum differences of the $R$ filter magnitudes between 
    seasons  and in $\sigma$ units for the stars
    ``R'', ``S'', ``T'', ``U'', and ``V'' in  Fig.~\ref{fig:full}.}
  \label{tab:refgal}
  \centering
  \begin{tabular}{ccc}
    \hline\hline
    star  & $\Delta m_R$ & ${\Delta m_R}/{\sigma_R}$ \\
    \hline
    ``R'' &    0.014     & 0.67 \\
    ``S'' &    0.030     & 1.29 \\
    ``T'' &    0.058     & 2.69 \\
    ``U'' &    0.037     & 0.84 \\
    ``V'' &    0.020     & 0.90 \\
    \hline
  \end{tabular}
\end{table}
% =====================================

We monitored \um during four seasons (2008--2011) using the Danish
$1.54\meter$ telescope at the La Silla Observatory, equipped with the
DFOSC instrument, providing $2147\times 2101$ pixel CCD frames over a
field of view of $13.7\arcmin\times13.7\arcmin$ with a declared
resolution of $0.39\arcsec/\rm pixel$. The RON (read-out-noise) of the
CCD camera in high-mode (gain $g=0.74$ electron/ADU) is $3.1$
electrons per pixel.  With the exception of the re-aluminization of
the primary mirror in 2009, the configuration software/hardware of the
telescope did not change over the four seasons of observation.  The
data were collected in the Bessel $V$, Bessel $R$, and Gunn $i$
filters\footnote { More details are available at
  \url{http://www.eso.org/lasilla/telescopes/d1p5/misc/dfosc_filters.html}}.

% We obtained a total of 130 images during the 2008 season: 
% 42 in the $V$ filter,
% 45 in the $R$ filter, and
% 43 in the $i$ filter, covering 15 nights each.\\
% During the 2009 season, the amount of collected images was 95:
% 34 in the $V$ filter,                                            
% 35 in the $R$ filter, and                                        
% 26 in the $i$ filter, covering 13, 12 and 9 nights, respectively.\\
% Then, during the 2010 season the total amount of images was 150:
% 72 in the $V$ filter, and
% 78 in the $R$ filter, covering 23 and 26 nights, respectively.\\
% Finally, during the 2011 season we obtained 113 images:
% 51 in the $V$ filter,                                            
% 53 in the $R$ filter, and                                        
% 9  in the $i$ filter, covering 15, 16 and 1 nights, respectively.

We obtained a total number of 488 $VRi$ images corresponding to 160
nights over the four seasons. The details are given in
Table~\ref{tab:obs}.

In 2010, no $i$ filter image was taken, as the monitoring was foreseen
since the beginning in the only $VR$ filters, and the $i$ filter
images were taken depending on the remaining telescope time with
respect to the other MiNDSTEp parallel projects. All the frames were
acquired with a $180\second$ exposure.

We treated the images following the same procedure as those relative
to \he described in a previous paper \citep{ricci11}, with the
exception that we used the images already de-biased and flat-fielded
\emph{in loco} by the IDL (Interactive Data Language) automatic
pipeline used at the Danish Telescope for the daily monitoring of the
bulge microlenses.

%%%%%%%%%%%%%%%%%%%%%%%%%%%%%%%%%%%%%%%%%%%%%%%%%%%%%%%%%%%%%%%%%%%%%
\section{Data reduction}
\label{sec:red}
% -------------------------------------------------------
\begin{figure}[tbp]
%  \centering \includegraphics[height=5.0cm]{galaxy-residuals.eps}
  \centering \includegraphics[width=8.7cm]{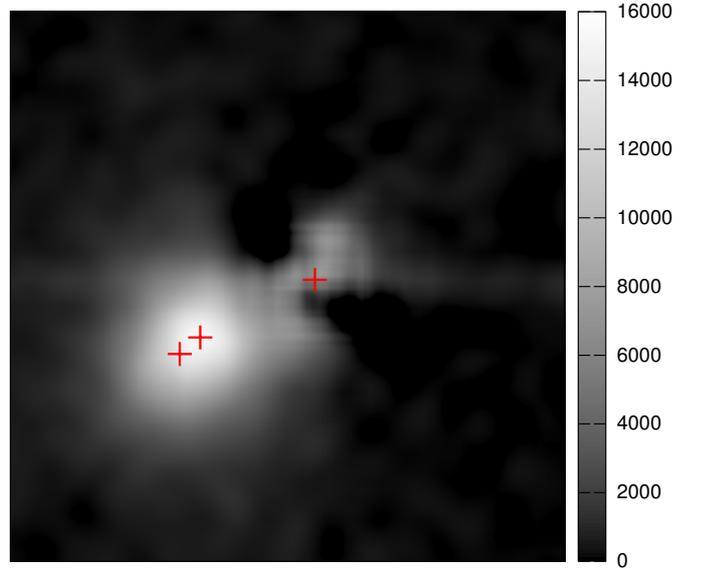}
  \caption{Reconstructed image of the lens galaxy of UM673.
      Orientation, pixel scale and marks are the same as those shown
      in Fig.~\ref{fig:galaxy-lens}. See the text for the details
      relative to the reconstruction technique. }
  \label{fig:galaxy-residuals}
\end{figure}
% ------------------------------------------------------

We carefully checked the history of the scale of the images between
the various seasons, and we found a constant value of
$0.395\arcsec/\rm pixel$. We froze this angular scale in the data reduction. We
also checked the evolution of the position angle between the CCD pixel
grid and the equatorial coordinate system, finding a change in angle
between the seasons: $4.5\arcmin$ between 2008 and 2009, $5.2\arcmin$
between 2008 and 2010, and $4.7\arcmin$ between 2008 and 2011. We took
this effect into account in our data reduction.

Finally, we checked the seeing values for all the observations. We
decided to fit the ``U'' star (see Fig.~\ref{fig:full}) with a
two-dimensional Gaussian function, and we found that the $R$ filter
images had the best seeing. We then decided to search for a
  suitable reference star in that filter.

We disregarded all those images for which the two lensed
  components were unresolved (seeing $>2\arcsec$).  Independently we
  measured the flux ratio between the two bright galaxies ``G1'' and
  ``G2'' (see Fig.~\ref{fig:full}) using aperture photometry (we
  integrated a square area of $40\times 40$ pixels centered on each
  galaxy). In the analysis we only used those images for which this
  flux ratio was stable, corresponding to a total of 9--18 images per
  season, depending on the filter.

The reference candidates are the stars ``R'', ``S'', ``T'', ``U'' and
``V'' in Fig.~\ref{fig:full}: we compared the fluxes of these stars
with the total flux of the bright galaxies ``G1'' and ``G2'' using
aperture photometry.  For this test we decided to use galaxies because
we can be sure of their stability.  Table~\ref{tab:refgal} contains
the maximum differences of the magnitudes between seasons and in sigma
units for the five concerned stars.

On the basis of this analysis, we conclude that star ``R'' and star
``V'' are comparably stable.  However, star ``V'' is closer to the
lens system, and it is therefore better to use its shape as a reference PSF
for the lens fitting.  Also, it had been found to be photometrically
stable by \cite{sinachopoulos01a} and \cite{nakos03}; finally it was
already used by \cite{nakos05} as a reference for the PSF fitting of
UM673.

From all these considerations, we decided to use star ``V'' as the
reference star for the PSF fitting of the lens system. To calibrate
the magnitudes in the $VRi$ filters, we used the values of the star
``V'' provided by \cite{nakos03}: $m_V=16.54\pm0.01$,
$m_R=16.00\pm0.01$, and $m_I=15.55\pm0.01$.  Moreover, we calculated
the $R$ magnitude of the ``G1'' and ``G2'' galaxies with aperture
photometry, using ``V'' as the reference star. We found values of
$m_R=17.47\pm0.03$ for ``G1'', and $m_R=17.92\pm0.05$ for ``G2''.

We tested if it was possible, on the basis of our data, to measure
independently the magnitude of the lens galaxy ``G'' in each band.  We
found that the $R$ band images had better quality, and we proceeded
using these images.  Each image was interpolated with a bicubic spline
and every pixel was divided in a grid of $20\times 20$ new sub-pixels.
Then we superposed these oversampled images and we summed them up to
obtain an oversampled image with a high signal-to-noise ratio (see
Fig.~\ref{fig:galaxy-lens}).  We used the ``V'' reference star as
reference PSF.  We fitted the gravitational lens system with two PSFs
for the ``A'' and ``B'' lensed components, fixing their relative
astrometry.  We then adjusted the scale factors of those two PSFs to
retrieve the uncontaminated image of the background lens galaxy. We
used aperture photometry to derive its magnitude relatively to the
``V'' reference star.

  To improve the accuracy of the photometry, we added two
  factors to scale the fluxes of the ``A'' and ``B'' lensed
  components.  We varied the factor of the ``A'' component
  from $0.94$ to $1.1$ with a step of $0.0022$, and we varied the
  factor of the ``B'' component from $0.2$ to $1.1$ with a
  step of $0.04$. 

First we constructed an array of residual maps for these two factor
combinations, and for each residual map we calculated the coordinates
of the light center of the galaxy ``G''.  As a criterion for the
correctness of the obtained galaxy image we chose the distance from
its light center to the expected one, provided by the accurate
astrometry measurements. The distance between the ``B'' lensed
component and the galaxy ``G'' provided by HST data is $0.38\arcsec$,
which is $\approx 20$ new sub-pixels. So we assumed that the distance
between the obtained and expected light center of ``G'' should be less
than half the distance between the galaxy ``G'' and the ``B'' lensed
component ($<10$ new sub-pixels).

We applied the same criterion between the expected and
  observed position of the maximum of light of the galaxy
  ``G''. Indeed, the light center of ``G'' may be slightly offest from
  its maximum of light. 

We considered that the overlap between the regions where these
  two conditions are satisfied fixes the region of allowed values for
  the two scale factors, and the minimum and maximum magnitudes of
  galaxy ``G'' which are $19.02$ and $19.56$, respectively.  From
  that, we then independently calculated the magnitude of the lensing
  galaxy ``G'' in the $R$ band as $19.29\pm0.27$.  If we calculate
  this value as an average magnitude over all the allowed values for
  the two scale factors, we obtain $19.27\pm0.15$.  Both these values,
  within the error bars, are in good agreement with the HST data. An image
  of the reconstructed galaxy is shown in
  Fig.~\ref{fig:galaxy-residuals}.  

Therefore, in the following analysis we considered the magnitudes of
the lens galaxy ``G'' as being those previously measured with HST.
HST results (named $G_{HST}$), obtained using HST filters, were
converted to the ground-based photometric system by \cite{lehar00}
using \cite{holtzman95} calibrations.  The $V$, $R$ and $i$ magnitudes
that they derived for the galaxy are: $G_V=20.81\pm0.02$,
$G_R=19.35\pm0.01$, and $G_I=18.72\pm0.03$.

%%%%%%%% Tables updated after Andrii calculations due to the SECOND
%%%%%%%% revision of the referee: THE VALUES ARE NORMALIZED TO THE
%%%%%%%% APERTURE PHOTOMETRY! REFERENCE DIRECTORY: "SUPERFINAL".

% =====================================
\begin{table}[bp]
  \caption{Average magnitudes for the gravitationally lensed 
    components of \um in the $VRi$ bands.}
\label{tab:ave}
  \centering
  \begin{tabular}{cccccccccc}
\hline\hline
     component       & season &      $V$       &      $R$         & $i$      \\           
\hline                                                                                  
 \multirow{4}{*}{A}  &  2008  & $16.79\pm0.02$  & $16.48\pm0.03$  & $16.27\pm0.03$ \\   
                     &  2009  & $16.88\pm0.04$  & $16.59\pm0.04$  & $16.32\pm0.03$ \\   
                     &  2010  & $16.84\pm0.01$  & $16.55\pm0.02$  &                \\  
                     &  2011  & $16.77\pm0.03$  & $16.48\pm0.03$  & $16.22\pm0.01$ \\   
\hline                                                                             
 \multirow{4}{*}{B}  &  2008  & $19.13\pm0.06$  & $19.09\pm0.05$  & $18.80\pm0.10$ \\  
                     &  2009  & $19.20\pm0.04$  & $19.16\pm0.05$  & $18.85\pm0.04$ \\  
                     &  2010  & $19.16\pm0.06$  & $19.10\pm0.05$  &                \\ 
                     &  2011  & $19.18\pm0.05$  & $19.22\pm0.10$  & $18.84\pm0.06$ \\  
\hline                                                                             
\multirow{4}{*}{B+G} &  2008  & $18.92\pm0.07$  & $18.46\pm0.05$  & $18.01\pm0.10$ \\   
                     &  2009  & $18.98\pm0.04$  & $18.50\pm0.05$  & $18.03\pm0.05$ \\    
                     &  2010  & $18.94\pm0.06$  & $18.47\pm0.05$  &                \\    
                     &  2011  & $18.96\pm0.06$  & $18.53\pm0.10$  & $18.03\pm0.07$ \\   
\hline
\end{tabular}
\end{table}
% =====================================

We applied the PSF fitting technique while accounting for the magnitude of
the lens on the best frames previously chosen, by refining the robust
method already used in our previous work \citep{ricci11}.

Our method is based on the simultaneous fit of each
    frame with two PSFs for the ``A'' and ``B'' components, and the de
    Vaucouleurs profile for the lens galaxy ``G'', fixing the relative
    astrometry between the components in accordance with measurements
    from \cite{castles}.  We also fixed the magnitudes of the galaxy
    ``G'' to the above mentioned values.  For a more accurate fitting,
    we used a bicubic interpolation of the images. 

\cite{koptelova10, koptelova12} observed \um with the
    $VRI$ filters, and derived the photometry without separately
    taking into account the magnitude of ``B'' and that of the lens galaxy
    ``G''.  As the lens galaxy is located very close to the ``B''
    component, and for  comparison with other works, we also
    calculated the magnitudes of the ``B''+``G'' components as a
    simple superposition of their fluxes.  Let us label ``B+G'' the
    results obtained in this way.

%%%%%%%%%%%%%%%%%%%%%%%%%%%%%%%%%%%%%%%%%%%%%%%%%%%%%%%%%%%%%%%%%%%%% 
\section{Results}
\label{sec:res}
% -------------------------------------------------------
\begin{figure}[tbp]
  \centering \includegraphics[width=8.7cm]{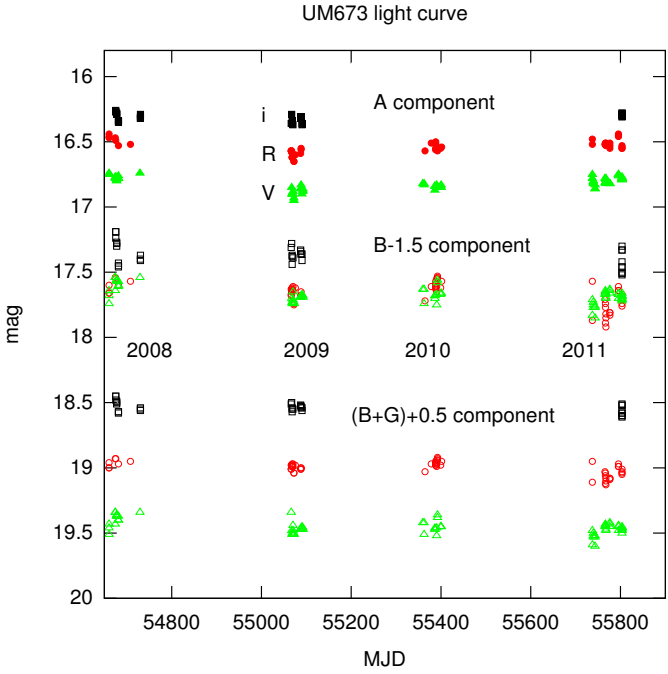}
  \caption{Light curves in the $VRi$ filters of the lensed components
    ``A'' and ``B'' of the gravitationally lensed quasar \um. The plot
    also shows the ``B+G'' values.
% , obtained by fitting the ``B''
%     component which includes the contribution due to the lens galaxy
%     ``G''. 
    The ``B'' and ``B+G'' light curves have been
      shifted by $-1.5$ and $0.5$ mag, respectively. Typical errors of
      individual observations are near $0.02$ and $0.05$--$0.08$ mag
      for the ``A'' and ``B'' components, respectively.
}
  \label{fig:curves}
\end{figure}
% -------------------------------------------------------
% -------------------------------------------------------
\begin{figure}[tbp]
  \centering \includegraphics[width=8.7cm]{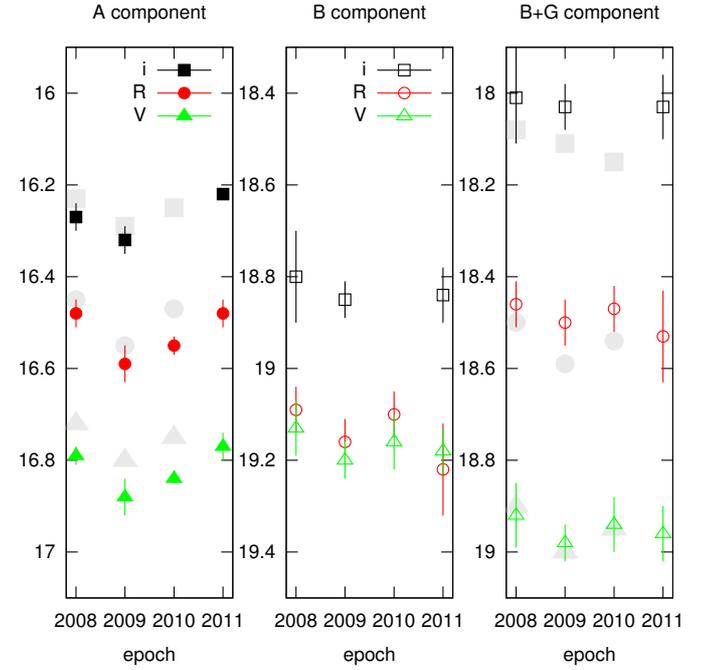}
  \caption{Average light curves over the four seasons of observation
    for the two lensed components ``A'' and ``B''. The ``B+G'' average
    light curve is also shown (see the text for details). The error
    bars indicate the standard deviation over the epoch.
    The larger background symbols show recent results
        independently obtained by \cite{koptelova12}. }
  \label{fig:medvar}
\end{figure}
% -------------------------------------------------------

%%%%%%%%%%%%%%%%%%%%%%%%%%%%%%%%%%%%%%%%%%%%%%%%%%%%%%%%%%%%%%%%%%%%% 
\subsection{Flux variations}
% =====================================
\begin{table}[bp]
  \caption{Average $R-i$ and $V-R$ color indices for the gravitationally lensed   components of \um.}
\label{tab:col}
  \centering
  \begin{tabular}{ccccc}
\hline\hline
     component       & season &       $R-i$       &      $V-R$       & $V-i$      \\
\hline
 \multirow{4}{*}{A}  &  2008  & $  0.21\pm0.04 $  & $  0.32\pm0.04 $ & $   0.52\pm0.03   $ \\
                     &  2009  & $  0.26\pm0.05 $  & $  0.29\pm0.06 $ & $   0.56\pm0.05   $ \\
                     &  2010  &                   & $  0.29\pm0.02 $ & \\      
                     &  2011  & $  0.26\pm0.03 $  & $  0.30\pm0.04 $ & $   0.55\pm0.03   $ \\  
\hline
 \multirow{4}{*}{B}  &  2008  & $  0.28\pm0.11 $  & $  0.04\pm0.08 $ & $   0.32\pm0.12   $ \\
                     &  2009  & $  0.32\pm0.06 $  & $  0.04\pm0.06 $ & $   0.35\pm0.06  $ \\
                     &  2010  &                   & $  0.06\pm0.07 $ & \\
                     &  2011  & $  0.38\pm0.11 $  & $ -0.04\pm0.11 $ & $   0.34\pm0.08  $ \\
\hline
\multirow{4}{*}{B+G} &  2008  & $  0.45\pm0.11 $  & $  0.46\pm0.08 $ & $   0.91\pm0.12   $ \\
                     &  2009  & $  0.47\pm0.07 $  & $  0.48\pm0.06 $ & $   0.95\pm0.07   $ \\
                     &  2010  &                   & $  0.48\pm0.08 $ & \\
                     &  2011  & $  0.51\pm0.12 $  & $  0.43\pm0.11 $ & $   0.93\pm0.09   $ \\
\hline                 
\end{tabular}
\end{table}
% =====================================

The separate light curves of the two lensed components ``A'' and ``B''
of \um and the ``B+G'' light curve are shown in Fig.~\ref{fig:curves}.

For a robust measurement of variability, we calculated the average and
the standard deviation over each season. Then, we
    also measured the photometry  of the whole system
    $(A+B+G)_{aperture}$ using aperture photometry.  The aperture
    photometry was calculated using two independent routines: a custom
    routine set up by our team, and the IRAF \texttt{daophot} package.
    As the results were robust and coherent between each other, we
    decided to normalize the averaged PSF fitting results to aperture
    photometry.  We then calculated for each year a normalization
    parameter $k= [ (A+B+G)_{aperture} - G_{HST} ] /
    (A_{PSF}+B_{PSF})$, and we corrected all PSF magnitudes for
  $k$.  These averaged results are in agreement with the non
normalized results, and are shown in Fig.~\ref{fig:medvar} and in
Table~\ref{tab:ave}.

Furthermore, in Table~\ref{tab:ave} we list the
    magnitudes of ``B+G'' as a mere superposition of
    their fluxes. The contribution due to the galaxy ``G'' in the total
    flux of the unresolved component ``B+G'' is quite important: near
    18\%, 45\% and 52\%  in the $V$, $R$ and $i$ bands, respectively.

We see an initial common behavior for the different filters and
components: the flux slightly decreases between the 2008 and  2009
seasons, and increases between the 2009 and  2010 seasons. Then,
during the 2011 season, the flux of the ``A'' component keeps
increasing, while the ``B'' component slightly decreases.

In particular, in the $V$ filter we notice a decrease in flux by
$0.09$ mag between the 2008 and 2009 seasons for the ``A''
component (corresponding to a decrease of $3\sigma$), and an increase
in flux by $0.10$ mag between the two successive seasons
(2009--2011).  The flux of the ``B'' lensed component, as well as of
``B+G'', slightly decreases in this filter over the four seasons, but
not significantly.

In the $R$ filter the behavior is the same: for the ``A'' component
the flux decreases by $0.11$ mag (above $3\sigma$) between
the first two seasons and successively increases by $0.11$ mag between
the 2009 and 2011 seasons.  The flux of the ``B'' lensed component
slightly decreases, as well as the flux of ``B+G'', with a less
significant amplitude.

Finally, in the $i$ filter we notice less evident trends than detected
in the other filters, excepted for the brighter ``A'' lensed
  component which presents a smaller decrease in flux between the
  first two seasons and a further increase by $0.10$ mag between 2009
  and 2011.

Our results are in good agreement with \cite{koptelova12}
  recent results for the same epochs (see the larger background
  symbols in Fig.~\ref{fig:medvar}).  We obtain for the ``A'' lensed
  component a magnitude $\approx0.02$--$0.08$ larger for all the
  filters.  The magnitudes of ``B+G'' are slightly smaller: within
  $2\sigma$ in the $R$ and $i$ bands. These differences might derive
  from using different techniques for PSF fitting and/or setting the
  photometric zero-points.

%%%%%%%%%%%%%%%%%%%%%%%%%%%%%%%%%%%%%%%%%%%%%%%%%%%%%%%%%%%%%%%%%%%%% 
\subsection{Color variations}
% -------------------------------------------------------
\begin{figure}[tbp]
  \centering \includegraphics[width=8.3cm]{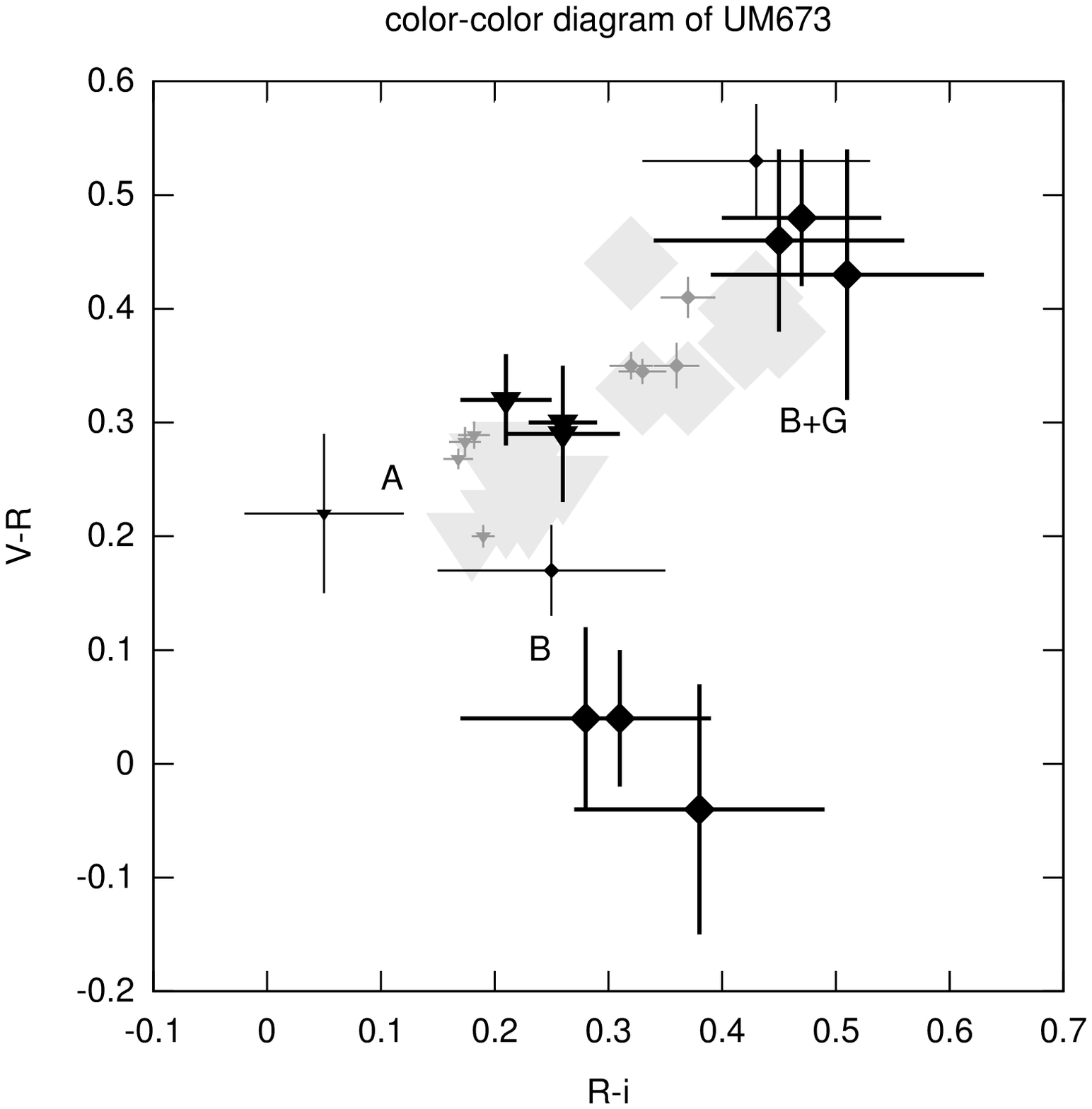}
  \caption{Color-color diagram for the 2008, 2009 and 2011 seasons
    (black bold dots) of the two lensed components ``A'' and ``B''
    of \um. The ``B+G'' values are relative to the color indices of
    the ``B'' component that includes the contribution of the lens
    galaxy, as in the approach of \cite{koptelova10}.  The diagram
    also includes HST \citep{castles} data (black light points) and
    \cite{koptelova10} data (little gray points).  The
        larger background symbols refer to the data from
        \cite{koptelova12}.}  
  \label{fig:color}
\end{figure}
% -------------------------------------------------------

From the data collected during the 2008, 2009, and 2011 seasons, we
were able to build a color-color diagram to search for color
variations of the two lensed components and of ``B+G'' with time. The
results are shown in Fig.~\ref{fig:color} and in Table~\ref{tab:col}.

All color variations over each epoch are found to be within the
error bars. Our results also show that within these error bars the
color indices of the ``A'' component and of ``B+G'' are coherent with
the work of \cite{koptelova10, koptelova12} data, and we find small
variations with respect to HST data which are relative to 1994.

% -------------------------------------------------------
\begin{figure}[tbp]
  \centering \includegraphics[width=8.7cm]{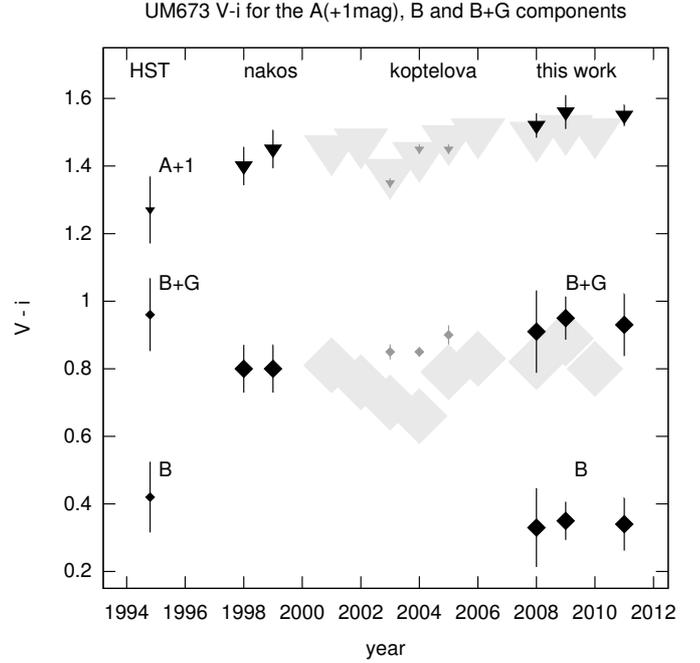}
  \caption{Evolution of the $V-i$ color index of ``A''
      (triangles), ``B'' and ``B+G'' (rhombi) with time, including
    the data from HST \citep{castles}, from \cite{nakos05}, from
    \cite{koptelova10} and from the present work.  Adaptation of
    recently published data by \cite{koptelova12} is also shown
    (larger background symbols). The ``A'' component is
      shifted by $+1$ mag.}
  \label{fig:tests}
\end{figure}
% -------------------------------------------------------

Moreover, the temporal evolution of the color index between the
observations of \cite{koptelova10, koptelova12} and the current data
%(see the arrows in Fig.~\ref{fig:color}) 
shows a weak trend indicating that the quasar becomes redder
  as its flux decreases, as already observed in our multi-color study
of the gravitationally lensed quasar \he \citep{ricci11}.

Galaxy ``G'' affects quite strongly the color of
    ``B+G''. We find a difference of $0.17$--$0.13$ mag between the
    $R-i$ color index of ``B+G'' and ``B'', a difference of
    $0.42$--$0.47$ mag in $V-R$ and of $0.59$--$0.60$ mag in $V-i$.  On the
    basis of HST data, corresponding differences in colors between the
    ``B+G'' and ``B'' components are $0.17$, $0.37$ and $0.54$
    \citep{lehar00}.  This supports the view that the
    contribution of galaxy ``G'' cannot be neglected in any
    considered band.

\cite{nakos05} reduced \um data in the $V$ and $i$
    filters by using three different techniques: MCS deconvolution,
    difference imaging, and PSF fitting.  The first two techniques
    allow in principle to get rid of the contribution of ``G'', while
    results obtained with PSF fitting are contaminated by the lens
    galaxy.  Despite this, \cite{nakos05} results are coherent with
    each other.  We compared their $V-i$ color index obtained by the
    different methods.  \cite{nakos05} obtained differences on the
    $V-i$ color index between ``B+G'' and the ``B'' components smaller
    than $0.04$ mag, which is comparable with their photometric
    errors.  This is in contradiction with \cite{lehar00} results and
    our results, which lead to $0.54$ and $0.59-0.60$ mag,
    respectively.  As it was shown above, the brightness of galaxy
    ``G'' cannot be neglected in the $V$, $R$ or $i$ bands.
    The contribution of galaxy ``G'' significantly changes the color of
    ``B+G''.

In Fig.~\ref{fig:tests} we compare the evolution of the
    $V-i$ color index with time, by using the data collected from HST
    \citep{castles}, \cite{nakos05}, and
    \cite{koptelova10}.  An adaptation of recently published data by
    \cite{koptelova12} is also shown.  We find good agreement with
    their data.

%%%%%%%%%%%%%%%%%%%%%%%%%%%%%%
\subsection{``Global $i$'' light curve}
\label{globcurv}
% -------------------------------------------------------
\begin{figure}[tbp]
  \centering \includegraphics[width=8.7cm]{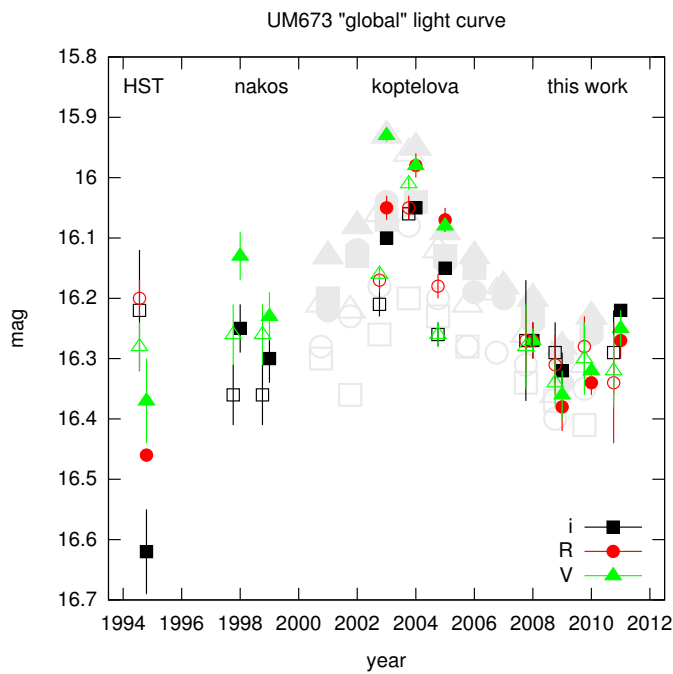}
  \caption{``Global $i$'' light curve of \um built by
      including data from HST \citep{castles}, \cite{nakos05},
      \cite{koptelova10} and the present work. An adaptation of
      recently published data by \cite{koptelova12} is also shown
      (larger background symbols). The technique used to build this
      curve is explained in detail in Sect.~\ref{globcurv}. In
      particular, filled and open symbols are used for the ``A'' and the ``B+G''
      lensed components, respectively. 
}
  \label{fig:time}
\end{figure}
% -------------------------------------------------------

A ``global $i$'' light curve which also includes the results of
\cite{castles}, \cite{nakos05} and \cite{koptelova10, koptelova12} is
shown in Fig.~\ref{fig:time}. To construct this figure, first we
shifted in time the light curve of  ``B+G'' using the
value of the time delay (89 days) provided by
    \cite{koptelova12}.  Then we calculated for each filter the
average 2008 difference in magnitude between the two components, and
we corrected the ``B+G'' light curve for these values. Finally, we
corrected the $V$ and $R$ light curves of both components by their
average 2008 $V-i$ and $R-i$ color indices, respectively. We chose the
2008 season as a reference only because it represents the beginning of
our observations.  Fig.~\ref{fig:time} shows that the flux of the
quasar intrinsically varied over the different seasons, with an
amplitude of $\approx 0.6$ mag, peak-to-valley over the last
two decades.

%%%%%%%%%%%%%%%%%%%%%%%%%%%%%%%%%%%%%%%%%%%%%%%%%%%%%%%%%%%%%%%%%%%%% 
\section{Conclusions}
\label{sec:conc}

We have presented a photometric monitoring, carried out during four epochs
in three different filters, of the doubly imaged quasar \um.

The results show a significant decrease in flux of the ``A'' lensed
component between the first two seasons (2008--2009), and a smaller
increase between the successive three seasons (2009--2011). This
behavior is mostly significant  in the $V$ and $R$ bands.

Moreover, our observations are in good agreement with the previous
works carried out by \cite{castles}, \cite{koptelova10}, and
\cite{koptelova12} in terms of flux variations and color index.  We
also separated the contribution of the lens galaxy from the fainter
lensed component, showing the effects of this operation on the color
index of the latter.  We conclude that the contribution of the lens
galaxy in the photometry of \um cannot be neglected and we give an
independent estimation of the magnitude of the lens galaxy.

Further observations could help in corroborating the separate color
variations of the components, and the slight flux trend observed
between the seasons.

%\newpage

%%%%%%%%%%%%%%%%%%%%%%%%%%%%%%%%%%%%%%%%%%%%%%%%%%%%%%%%%%%%%%%%%%%%
\begin{acknowledgements}
  This research was supported by ARC -- Action de recherche
  concert\'ee (Communaut\'e Fran\c caise de Belgique -- Acad\'emie
  Wallonie-Europe). DR (boursier FRIA) acknlowledges GLObal Robotic
  telescopes Intelligent Array for e-Science (GLORIA), a project
  funded by the European Union Seventh Framework Programme
  (FP7/2007-2012) under grant agreement number 283783.  AE is the
  beneficiary of a fellowship granted by the Belgian Federal Science
  Policy Office. AE is also grateful for partial support in the
  framework of the NASU Target Program ``CosmoMicroPhysics''. NK
  received funding from the European Community's Seventh Framework
  Programme (/FP7/2007-2013/) under grant agreement No 229517. MD, MH
  and CL acknowledge the Qatar Foundation for support from QNRF grant
  NPRP-09-476-1-078. Operation of the Danish $1.54\meter$ telescope is
  supported by the Danish National Science Research Council (FNU).

\end{acknowledgements}

\bibliography{biblio} % your references biblio.bib
\bibliographystyle{aa} % style aa.bst

\end{document}